\documentclass[sigconf]{acmart}
\AtBeginDocument{%
  \providecommand\BibTeX{{%
    \normalfont B\kern-0.5em{\scshape i\kern-0.25em b}\kern-0.8em\TeX}}}

\setcopyright{acmcopyright}
\copyrightyear{2021}
\acmYear{2021}
\acmDOI{10.1145/3475959.3485394}

\copyrightyear{2021}
\acmYear{2021}
\setcopyright{rightsretained}
\acmConference[MuCAI '21]{Proceedings of the 2nd ACM Multimedia Workshop on Multimodal Conversational AI}{October 24, 2021}{Virtual Event, China}
\acmBooktitle{Proceedings of the 2nd ACM Multimedia Workshop on Multimodal Conversational AI (MuCAI '21), October 24, 2021, Virtual Event, China}\acmDOI{10.1145/3475959.3485394}
\acmISBN{978-1-4503-8679-1/21/10}

\begin{document}
 \fancyhead{} 
\title{Towards a Real-time Measure of the Perception of Anthropomorphism in Human-robot Interaction}


\author{Maria Tsfasman}
\email{M.Tsfasman@tudelft.nl}
\authornote{All three authors contributed equally to this research.}
\affiliation{%
  \institution{Delft University of Technology}
  \streetaddress{Van Mourik Broekmanweg 6}
  \city{Delft}
  \country{The Netherlands}
  \postcode{2628 XE}
}

\author{Avinash Saravanan}
\email{A.Saravanan@student.tudelft.nl}
\authornotemark[1]
\affiliation{%
  \institution{Delft University of Technology}
  \streetaddress{Mekelweg 5}
  \city{Delft}
  \country{The Netherlands}
  \postcode{2628 CD}
}

\author{Dekel Viner}
\email{D.Viner@student.tudelft.nl}
\authornotemark[1]
\affiliation{%
  \institution{Delft University of Technology}
  \streetaddress{Mekelweg 5}
  \city{Delft}
  \country{The Netherlands}
  \postcode{2628 CD}
}

\author{Daan Goslinga}
\email{D.B.Goslinga@student.tudelft.nl}
\affiliation{%
  \institution{Delft University of Technology}
  \streetaddress{Mekelweg 5}
  \city{Delft}
  \country{The Netherlands}
  \postcode{2628 CD}
}

\author{Sarah de Wolf}
\email{S.C.M.deWolf@student.tudelft.nl}
\affiliation{%
  \institution{Delft University of Technology}
  \streetaddress{Mekelweg 5}
  \city{Delft}
  \country{The Netherlands}
  \postcode{2628 CD}
}

\author{Chirag Raman}
\email{C.A.Raman@tudelft.nl}
\affiliation{%
  \institution{Delft University of Technology}
  \streetaddress{Mekelweg 5}
  \city{Delft}
  \country{The Netherlands}
  \postcode{2628 CD}
}

\author{Catholijn M. Jonker}
\email{C.M.Jonker@tudelft.nl}
\affiliation{%
  \institution{Delft University of Technology}
  \streetaddress{Van Mourik Broekmanweg 6}
  \city{Delft}
  \country{The Netherlands}
  \postcode{2628 XE}
}

\author{Catharine Oertel}
\email{C.R.M.M.Oertel@tudelft.nl}
\affiliation{%
  \institution{Delft University of Technology}
  \streetaddress{Van Mourik Broekmanweg 6}
  \city{Delft}
  \country{The Netherlands}
  \postcode{2628 XE}
}

\renewcommand{\shortauthors}{Tsfasman, Saravanan and Viner, et al.}

\begin{abstract}
How human-like do conversational robots need to look to enable long-term human-robot conversation? One essential aspect of long-term interaction is a human's ability to adapt to the varying degrees of a conversational partner's engagement and emotions. Prosodically, this can be achieved through (dis)entrainment. While speech-synthesis has been a limiting factor for many years, restrictions in this regard are increasingly mitigated.
These advancements now emphasise the importance of studying the effect of robot embodiment on human entrainment. In this study, we conducted a between-subjects online human-robot interaction experiment in an educational use-case scenario where a tutor was either embodied through a human or a robot face. 43 English-speaking participants took part in the study for whom we analysed the degree of acoustic-prosodic entrainment to the human or robot face, respectively. We found that the degree of subjective and objective perception of anthropomorphism positively correlates with acoustic-prosodic entrainment.

\end{abstract}

\ccsdesc[300]{Human-centered computing~Empirical studies in collaborative and social computing}
\ccsdesc[500]{Human-centered computing~Empirical studies in interaction design}
\ccsdesc{Computer systems organization~Robotics}

\keywords{multi-modal, human-robot interaction, prosody, acoustic-prosodic entrainment}

\maketitle

\section{Introduction}
To what degree the anthropomorphism of robots affects human-robot conversation is becoming an increasingly pressing question.
Underlying technology such as speech synthesis, speech recognition, and natural language understanding has improved recently to such a degree that long-term human-robot applications are becoming much more feasible.
A critical aspect of long-term human-robot interaction is the creation of rapport, a social clicking or bonding between interaction partners \cite{TickleDegnen1990}.
In human-human interaction, rapport has been shown to be related to acoustic-prosodic entrainment \cite{Lubold2014EntrainmentRapport}, an ability of humans to adapt their prosody to each other within the conversation \cite{levitan-etal-2012-acoustic}. Acoustic-prosodic entrainment can also be a reliable indicator of conversational involvement and interaction quality \cite{de2011measuring, reichel2018prosodic}. Higher acoustic-prosodic entrainment can indicate more interest in the topic discussed, higher levels of attention \cite{LevitanBenusGalvez2016, reichel2018prosodic} and can also be related to the activation of mirror neurons \cite{Levy2016MirrorEntrainment, Gog2008MirrorNeurons}.

In human-robot interaction, acoustic-prosodic entrainment has been implemented as a tool to increase a robot's social presence and a user's rapport towards it \cite{Lubold2016RobotEntrainmentIncreasesRapportAndSocPresence}. The ability of the robot to entrain to the user has also been shown to increase children's engagement within the interaction \cite{Sadouohi2017RobotEntrainmentIncreasesSynchrony&Engagement}.

The famous 'uncanny valley' phenomenon \cite{mori_1970} implies that if the robot is too human-like, it can decrease the likeability of the robot. Yet, human-like appearance has shown to increase perceived trust-worthiness \cite{deVisser2016HumanlikenessAndTrust, Natarajan2020AnthropomorphismAndTrust} and social presence \cite{RanHee2014AnthropomSocialPresence, Schuetzler2020SocialPresenceAnthropomorphism} of conversational agents. Questions related to the link between anthropomorphism of and attitudes towards virtual agents are widely studied \cite{Fink2012AnthropomorphismReview}, yet questions related to how anthropomorphism influences the user's behaviour are less explored.

Certain qualities within Instructors in educational contexts can potentially impact the educational outcomes as well.
In an educational context, the anthropomorphism of the tutor has been shown to improve the understanding and memorisation of information, and human tutors still show better results than a robot or a tablet \cite{Westlund2015ACO}. Thomason et. al \cite{ThomasonNguyen} investigated how acoustic-prosodic entrainment correlates with the quality of information acquisition in a tutoring experiment.
They compared knowledge gain and acoustic-prosodic entrainment of students when learning from a tutor with a human in contrast to a synthesized voice. Based on amplitude and pitch features, \cite{ThomasonNguyen} show that higher acoustic-prosodic entrainment positively correlates with students’ knowledge gain. More importantly, they found that students' acoustic-prosodic entrainment was higher in the human voice scenario. In \cite{ThomasonNguyen}'s study, the virtual tutor was a voice assistant, and the conditions were different in the human-likeness of the virtual tutor's speech. But does visual human-likeness affect the entrainment in the same way? 

In this paper, we investigate the effect of a virtual tutor's human versus machine-like appearance on a user's prosodic entrainment. In fact, there have been human-robot interaction studies measuring user entrainment in human-robot interaction. \citet{Breazeal2001EntrainmentHRI} showed that humans are able to adapt their turn-taking behaviour to a robot. \citet{Strupka2016HumanEntrainmentHRI} investigated how humans adapt their speech to the robots of different genders found that participants exhibited speech divergence (the opposite of entrainment) in both conditions.

If there is a connection between anthropomorphism of robot appearance and acoustic-prosodic entrainment, that could indicate the importance of making virtual agents and social robots as human-like as possible for long-term interaction scenarios such as in a hybrid-intelligence scenario \cite{Akata2020}.

Real-time assessment of entrainment could act as a non-verbal indication of engagement and rapport towards the robot in the future. This way, no additional questionnaires are needed to access the user perception of the robot. It could be done in an online manner, by processing the user's speech.

Another field that could benefit from understanding the connection between anthropomorphism and acoustic-prosodic entrainment is multi-modal addressee detection.  There are many studies focusing on advancing automatic detection of human- versus system- addressed speech  \cite{Tsai2015addressee, shriberg2013addressee, shriberg2012addressee}. All of the cited algorithms show to benefit from using prosodic features as predictors of addressee tags. However, they have not used acoustic-prosodic entrainment as their feature, and in case our hypothesis is confirmed, it might aid the automatic detection of whether the user is addressing a machine or another human.

\section{Research Question}

The research question we aim to answer in the present paper is following:
Does the type of facial embodiment of a conversational agent influence the level of acoustic-prosodic entrainment of a person interacting with it?

To answer this question in the present work we investigate human acoustic-prosodic entrainment in two conditions: 
\begin{enumerate}
    \item \textbf{Human condition} (Human Face, Human Voice) - A lesson is taught with pre-recorded videos of a human tutor.
    \item \textbf{Robot condition} (Robotic Face, Human Voice) - The same lesson is taught by a virtual agent mimicking the exact face movements and using the audio of the human tutor from the first condition.
\end{enumerate}

\subsection{Hypothesis}

In relation to audio cues, it has been tested and confirmed in \cite{ThomasonNguyen} that the synthesized voice modulates less acoustic-prosodic entrainment than the human voice. However, it hasn't been tested in relation to visual cues independently of the phonetic or non-verbal cues. In this experiment, our main hypothesis is that the human appearance of a virtual tutor will incite more acoustic-prosodic entrainment than the robotic tutor with the same voice and facial expressions. Our second expectation is that the perceived anthropomorphism of the tutor will positively correlate with user acoustic-prosodic entrainment.

We hypothesise that :
\begin{enumerate}
    \item  H1: participants show greater entrainment towards the human than the robot face
    \item  H2: the greater the participant's perception of anthropomorphism the greater the degree of entrainment
\end{enumerate}

\section{Method}\label{Method}

\subsection{Stimulus Preparation}

\subsubsection{Task}
Because of the COVID-19 pandemic, all participant interactions were carried out via Zoom.
The participants followed three lessons on random topics (about meatball production, beer crafting and venomous species). Each lesson lasted around 10 minutes and consisted of multiple pre-recorded videos taught by a virtual tutor (robot or human depending on the condition). Each pre-recorded video lasted for about 2-10 seconds and ended in an open question inviting participants for interaction. The participants were invited to verbally interact with the tutor throughout the whole lesson but especially after the tutor asks a question. After the video, a pre-recorded idle state was played for as long as a participant was replying. The idle states consisted of common back-channels (nodding, smiling, etc.) also pre-recorded with the same virtual tutor. The length of idle states was controlled by the experimenter within each experiment to avoid interrupting participants. After finishing all the lessons participants had to complete a questionnaire on their perception of the interaction and the tutor.

\subsubsection{Conditions}
There were two conditions: (1) \textbf{human} and (2) \textbf{robot}. In the human condition, the lessons were recorded from a male English-speaking actor. The only difference between the human and the robot condition was the appearance of the tutor. The audio from the human tutor recording was used in both conditions, the content and the post-questionnaire was also the same across conditions. Since the conditions had the same content, the experiment had a between-subject design. 

For the robot condition, we used the Furhat SDK \cite{AlMoubayed2012Furhat}. Furhat is "a social robot with human-like expressions and advanced conversational artificial intelligence (AI) capabilities" \cite{Furhat}. For the recording of the robot condition stimuli, we used the Furhat simulation. 

To make facial expressions as similar as possible between the two conditions, we used code that resynthesised the human tutor's face movements in Furhat. The importance of this step can be illustrated by Breazeal's \cite{Kismet} findings on the significance of body posture, head tilt and facial expression for human-robot entrainment. The code used OpenCV python library \cite{opencv_library} to track mouth movements, smile and head position and rotation in the video of the human tutor. It then proceeded to convert the resulting facial movements and their durations to the Furhat implementation.

Figure \ref{fig:stimuli} shows a screenshot of videos shown to the participants in two different conditions, side-by-side, in the same moment of time:
\begin{figure}[h]
  \centering
  \includegraphics[width=1.0\linewidth]{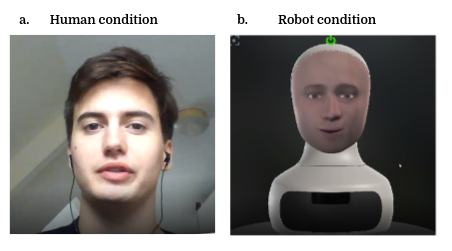}
  \caption{Screenshot of stimuli videos in human (a) and in robot (b) conditions, at the same point of time}
  \label{fig:stimuli}
\end{figure}


\subsubsection{Questionnaires}
After the participants have finished the virtual lessons, they answered questions relating to their experience of interacting with the tutor in a Qualtrics survey. These questions included Godspeed questionnaire scales on animacy and anthropomorphism \cite{bartneck2009measurement} and two questions on perception of whether the videos had smooth transitions between interaction and how interesting were the lessons.

\subsection{Experimental Set-Up.}

\subsubsection{Experimental Study}

We strove to reduce the differences between conditions to the level of human-likeness of tutor appearance to use conditions as an objective measure of anthropomorphism. For the subjective measure of anthropomorphism, we used the ratings collected from the Godspeed anthropomorphism scale. The subjective measure was meant to both validate the distinction between conditions and test our hypothesis. 

\subsubsection{Participants}
The experiment was conducted on 51 English-speaking participants, who have never met the tutor before and never came across Furhat robot. It was important to have participants that have never interacted with the human tutor since prior interactions might bias the way in which people display their entrainment \cite{Yuan2007}. Of these 51 participants, 43 have been included in the results. The 8 other participants have been excluded because of internet problems and recording errors, which made the data unusable. The resulting set of participants included 22 females and 21 males (mean age 28 +- 10). 

\subsubsection{Procedure}

Each experiment lasted for about 40 minutes altogether. Each participant was assigned a random condition. After signing a consent form for audio recording and data storage, participants had to join a Zoom call with the experimenter, and after being fully instructed, watched the lessons through experimenter's screen sharing. The experimenter's video was off throughout the whole call and the audio interaction was reduced to a minimum to avoid participants entraining with the experimenter.
The instructions for the participants were to wear a personal headset, listen carefully and verbally interact with the tutor as much as possible.

\subsection{Analysis}
Our experiment contains one independent variable (facial embodiment) and one dependent variable (acoustic-prosodic entrainment). Facial embodiment is a binary variable. Namely, the facial appearance is either a video of a human actor or a robotic face (Furhat robot). This distinction between conditions we use as a measure of objective anthropomorphism. The acoustic-prosodic entrainment is a variable that spans over multiple features and metrics extracted from the audio recording of participants' speech.

\subsubsection{Prosodic feature extraction}

To analyse the audio Parselmouth \cite{jadoul_2019} and Pydub \cite{robert_2011} python libraries were used. Parselmouth was used to extract pitch and RMS-intensity.

The prosodic features that we extracted are similar to the ones used in Levitan et al. \cite{LevitanGravanoWillson2012}: mean Intensity, max Intensity, mean Pitch, max Pitch. Because of the online setting of the experiment, we could not control for the quality of the microphone, most participants used wired earphones. This was done to avoid leaking of tutor voice into the audio recording of the participants and to reduce the noise captured in the recording. 

\subsubsection{Preprocessing of features}
 The prepossessing conducted on the data before computing the entrainment metrics included standardization and KNN regression.

In a normal conversation, there are many moments in which one person is silent while the other is speaking and vice versa. There are also moments in which both speakers are silent. In both of these scenarios, it would make no sense to compute values for the metrics for entrainment. We use KNN regression on the data in order to fill in those gaps \cite{galvez-etal-2020-unifying}. For every feature at each time point take we take the average of the k nearest values (k = 7 in our case). Where the distance per value is computed against the centre time point of the utterance. In other words $\frac{F_{start\_time} + F_{end\_time}}{2}$.

\subsubsection{Acoustic-prosodic entrainment metrics}
We used acoustic-prosodic entrainment metrics introduced by \cite{levitan-etal-2012-acoustic} and commonly used for measuring entrainment: proximity, convergence and synchrony.

\textbf{Proximity} is computed by taking the negative absolute difference per feature at every time point. In order to make our results comparable between participants of different voice characteristics (such as male and female voices) we standardized values to their z-score. 
\begin{equation}
  -|f^A(t) - f^B(t)|  
\end{equation}
Important to note that the metrics are computed after the KNN preprocessing so the time points are not referring to the raw data. The closer the metric value is to 0 the higher the assumed entrainment.

\textbf{Convergence} measures how proximity changes over time, where $D(t)$ stands for $-|f^A(t) - f^B(t)|$ 
\begin{equation}
    \frac{\int_{t_0}^{t_n}(D(t)-\bar{D}) * (t - \bar{t})dt}{\sqrt{\int_{t_0}^{t_n}(D-\bar{D})^2 dt\int_{t_0}^{t_n}(t-\bar{t})^2 dt}}
\end{equation}
Convergence applies Pearson correlation and a positive convergence for a feature suggests that over the course of the conversation the proximity between the tutor and the student increases. Meaning the feature values become more similar. Likewise, negative convergence for a given feature means it becomes more dissimilar.

\textbf{Synchrony} is here taken simply as Pearson correlation for a given feature between the tutor and student. 
\begin{equation}
    \frac{\int_{t_0}^{t_n}(f^A(t + \delta) - \bar{f^A}) * (f^B(t) - \bar{f^B})dt}{\sqrt{\int_{t_0}^{t_n}(f^A(t + \delta) - \bar{f^A})^2 dt * \int_{t_0}^{t_n}(f^B(t) - \bar{f^B})^2dt}}
\end{equation}

\subsubsection{Statistical Methods}
To investigate the differences between conditions, we utilized Kruskal-Wallis test. The reason for this is due to data not meeting the assumptions of Anova which are normality and homogeneity of variance. These assumptions were tested through Shapiro and Levene tests. Pearson correlation was computed to determine the significance and the direction of the correlations between the entrainment metrics and subjective measures of anthropomorphism, animacy and interest. Finally, power analyses were carried out to determine if the sample size and statistical power were appropriate and we found that the power was adequate (power > 0.8) for all mentioned significant results.

\section{Results}

\subsection{Experiment perception}
The experiment perception questions at the end of the experiment included animacy and anthropomorphism scores from Godspeed questionnaire in order to confirm the opposition of human vs machine in the conditions. They also included a question on how interesting participants found the lessons and how smooth the transitions between interactions were. The interest score was aimed to control for the fact that participants could be more interested and therefore more engaged in the lesson with Furhat because of its novelty. It also served as a measure of subjective engagement, since entrainment has been linked to engagement before \cite{Sadouohi2017RobotEntrainmentIncreasesSynchrony&Engagement}. The question on smoothness of transitions was aimed to control for the differences in conditions connected to the way they were merged together in the experimental stimuli. 

\begin{figure}[!h]
  \centering
  \includegraphics[width=1.0\linewidth]{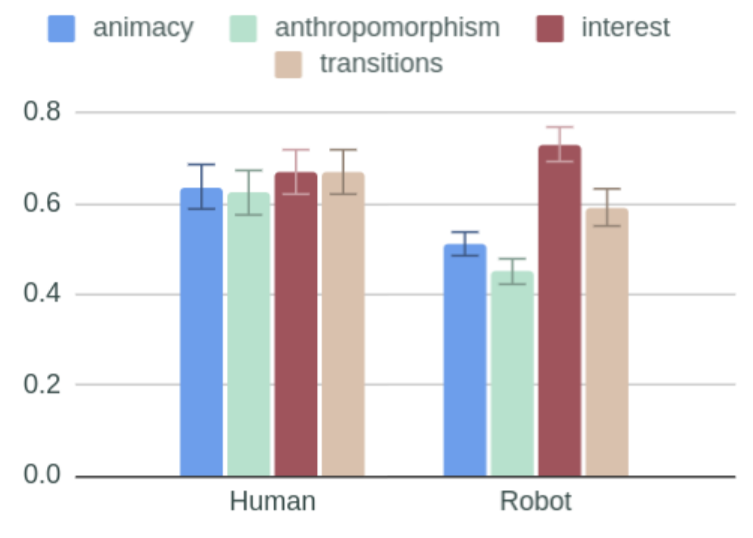}
  \caption{Average experiment perception scores for different conditions (error bars show standard error).}
  \label{fig:perception}
\end{figure}

Figure \ref{fig:perception} illustrates the differences between perception of each condition (human and Furhat). The y-axis shows the score normalised over the maximum score for each type of question, therefore for each perception parameter the maximum score is 1 and the minimum is 0.

The animacy and anthropomorphism scores were significantly higher in the human condition (p<0.01). This confirms our assumption of it being perceived more human-like. The animacy scores were also significantly higher in the human condition, as expected (p<0.01). The interest and the smoothness scores were insignificantly different between conditions, which confirms that the conditions were similar technically and content-wise.

\subsection{Acoustic-prosodic entrainment}

Our results indicate that, despite the online setting and artificial nature of the interactions, all participants entrained on the tutor voice: Every participant in both conditions had at least one feature with significantly positive convergence or at least one with significantly positive synchrony.

Although there was entrainment in both conditions, participants' proximity, convergence and synchrony in their intensity (mean and max) and max pitch were insignificantly different between conditions (p>0.01). 
However, for the mean pitch, which is a major predictor of acoustic-prosodic entrainment \cite{galvez-etal-2020-unifying}, the convergence was significantly lower in robot condition in comparison to the human condition (p<0.01). This said, proximity and synchrony by mean pitch by themselves were insignificantly different between conditions.

\begin{figure}[!h]
  \centering
  \includegraphics[width=0.7\linewidth]{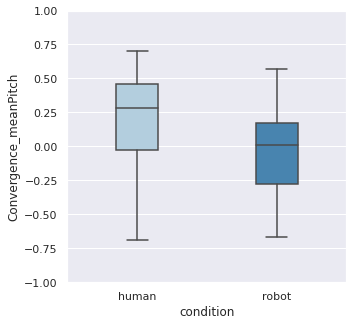}
  \caption{Differences in mean pitch convergence between conditions.}
  \label{fig:convergence_meanpitch}
\end{figure}

Although the variability between participants is quite high (see figure \ref{fig:convergence_meanpitch}), the convergence by mean pitch was significantly positive for 65\% of participants in human condition. In robot condition only 39 \% of participants displayed significantly positive convergence by mean pitch in robot condition. This means that the acoustic-prosodic entrainment was stronger in the human condition, confirming our hypothesis. 

\begin{figure}[!h]
  \centering
  \includegraphics[width=0.7\linewidth]{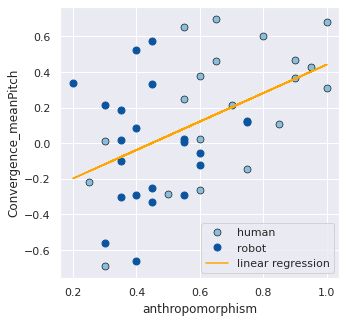}
  \caption{Significant positive correlation between convergence by mean pitch and perceived anthropomorphism.}
  \label{fig:convergence_meanpitch_anthrop}
\end{figure}

Not only objective human-likeness (i.e. distinction between condition), but also subjective perception of anthropomorphism appeared to positively correlate with convergence by mean pitch (Pearson correlation p<0.01). To illustrate the trend, we plotted linear regression over all participants figure \ref{fig:convergence_meanpitch_anthrop}.

\begin{figure}[!h]
  \centering
  \includegraphics[width=0.7\linewidth]{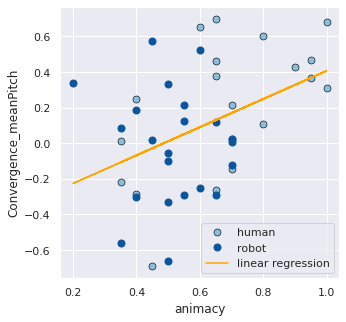}
  \caption{Significant positive correlation between convergence by mean pitch and perceived animacy.}
  \label{fig:convergence_meanpitch_animacy}
\end{figure}

Another significantly positive correlation can be noticed between convergence by mean pitch and the perception of animacy of the tutor (figure \ref{fig:convergence_meanpitch_animacy}).
The answers on a question on subjective engagement ('Rate how interesting did you find the lesson from 1 to 10') show similar trend: there was a significantly positive correlation of convergence by mean pitch to the interest score (Pearson correlation p<0.01).

\section{Discussion}

We found that mean pitch convergence was significantly higher in the human than in the robot condition. This confirms H1.
The fact that humans entrain more to other humans than to a robot is in line with findings by \citet{Strupka2016HumanEntrainmentHRI}.

We found that the acoustic-prosodic entrainment in mean pitch was positively correlated with participant's perception of human-likeness. In fact this was true for  both the subjective (perceived anthropomorphism) and objective (the distinction between conditions) anthropomorphism measures. This confirms H2.

The fact that we did not find prosodic entrainment in other prosodic features is in line with previous research on human-human entrainment \cite{galvez-etal-2020-unifying}. This might be related to the variability in participants' mother-tongue - although they all were English-speaking, they had different degree's of English proficiency and English accents; they also had different cultural backgrounds and varied in gender and age. Since all those can affect acoustic-prosodic entrainment \cite{levitan-etal-2012-acoustic, lewandowski2018}. A bigger sample study will be carried out in the future. 

Our results expand the previous study by \cite{ThomasonNguyen} to the visual domain: in \cite{ThomasonNguyen}'s experiment the less human-like voice triggered less acoustic-prosodic entrainment than a human voice. In our experiment, the robotic appearance of a tutor triggered less entrainment than a human tutor. This also is in line with the distinction between human-addressed and machine-addressed speech found in \cite{shriberg2013addressee, Tsai2015addressee}. Adding entrainment as one of the features could prove to be beneficial for those algorithms.

The fact that we discovered the effect of anthropomorphism on entrainment brings various implications, since there are many perceptual factors that are entailed by anthropomorphism in human-robot interaction and entrainment in human-human interaction. More human-like conversational agents seem more trust-worthy \cite{deVisser2016HumanlikenessAndTrust, Natarajan2020AnthropomorphismAndTrust} and socially present \cite{RanHee2014AnthropomSocialPresence, Schuetzler2020SocialPresenceAnthropomorphism}. Therefore, it may suggest that higher prosodic entrainment could correlate with conversational agent's perceived trustworthiness and social presence in human-robot interaction, as well as speakers' rapport \cite{Lubold2014EntrainmentRapport} and engagement \cite{levitan-etal-2012-acoustic, reichel2018prosodic}. 

All in all acoustic-prosodic entrainment appears to be a promising behavioural measure of access to a person's perception of animacy and anthropomorphism in his/her conversational partner. A real-time measure based on prosodic entrainment could widely benefit fields such as social robotics and hybrid intelligence \cite{Akata2020}. 
It might further be a relevant measure to further investigate in relation to human-human/machine addressee detection \cite{Tsai2015addressee, shriberg2013addressee,shriberg2012addressee}. 

On a cognitive level, there might be a link between anthropomorphism and entrainment via mirror neurons, which is activated when a human interacts or observes another human \cite{Rizzolatti2004Mirror, Gog2008MirrorNeurons}. The mirror neurons are thought to activate embodied experiences and therefore aid imitation learning \cite{Ramsey2021MirrorLearning}. Since entrainment is imitation in itself, mirror neurons can also be viewed as crucial mechanism in social entrainment \cite{Levy2016MirrorEntrainment}. This might explain why anthropomorphism of the tutor triggered more acoustic-prosodic entrainment in our experiment, and why interest and animacy scores also correlated with higher entrainment. 

\section{Conclusion and future work}
In this work, we investigated the effect of a human versus robot face on prosodic entrainment in an educational use-case scenario. 
We could show that humans converged to a higher degree in mean pitch to another human face than a robot face. Maybe more importantly, though, we could show that the greater the perception of animacy and anthropomorphism, the greater the degree of prosodic entrainment.
In future research, we plan to add the variable of age and gender to our experimental setup. Using purposeful manipulations of prosodic convergence, we aim to explore their effect on participants' recollection of the conversation.

\begin{acks}
  We would like to thank Navin Laxminarayanan Raj Prabhu for providing us with his code for replicating facial features in Furhat.
\end{acks}

\bibliographystyle{ACM-Reference-Format}
\bibliography{sample-base}

\end{document}